\let\orilabel\label
\documentclass[aps,pra,reprint,superscriptaddress,nofootinbib]{revtex4-2}
\let\label\orilabel

\usepackage{verbatim,amsmath,amssymb,color,appendix,algorithm,tensor,mathtools,mathrsfs,enumitem,graphicx,lipsum,xcolor}
\usepackage[final]{hyperref}
\usepackage[all]{hypcap}    
\definecolor{blue}{rgb}{0.0,0.0,0.8}
\hypersetup{
  colorlinks=true,
  linkcolor=blue,
  citecolor=blue,
  urlcolor=blue,
  linkbordercolor={0 0 1}
}

\newcommand{\bra}[1]{\langle{#1}|}
\newcommand{\ket}[1]{|{#1}\rangle}
\newcommand{\braket}[1]{\langle{#1}\rangle}

\begin{document}


\title{Efficient tensor network simulation for few-atom, multimode Dicke model via\\coupling matrix transformation}


\author{Christopher J. Ryu}
\affiliation{Department of Electrical and Computer Engineering, University of Illinois Urbana-Champaign, Urbana, IL 61801, USA}


\author{Dong-Yeop Na}
\affiliation{Department of Electrical Engineering, Pohang University of Science and Technology, Pohang 37673, Republic of Korea}

\author{Weng C. Chew}
\email{wcchew@purdue.edu}
\affiliation{Department of Electrical and Computer Engineering, University of Illinois Urbana-Champaign, Urbana, IL 61801, USA}
\affiliation{Elmore Family School of Electrical and Computer Engineering, Purdue University, West Lafayette, IN 47907, USA}

\author{Erhan Kudeki}
\affiliation{Department of Electrical and Computer Engineering, University of Illinois Urbana-Champaign, Urbana, IL 61801, USA}

\date{\today}

\begin{abstract}
We present a novel generalization of the chain mapping technique that applies to few-atom, multimode systems by making use of coupling matrix transformations. This is extremely useful for tensor network simulations of the multimode Dicke model and multi-spin-boson model because their coupling structures are altered from the star form to the chain form with near-neighbor interactions. Our approach produces an equivalent Hamiltonian with the latter coupling form, which we call the band Hamiltonian, and we demonstrate its equivalence to the multimode Dicke Hamiltonian. In the single atom case, our approach reduces to the chain mapping technique. When considering several tens of field modes, we have found that tensor network simulation of two atoms in the ultrastrong coupling regime is possible with our approach. We demonstrate this by considering a pair of entangled atoms confined in a cavity, interacting with thirty electromagnetic modes.
\end{abstract}


\maketitle

\section{Introduction}
The Dicke model describes the physics between a collection of two-level atoms and quantized electromagnetic field \cite{Dicke1954}. It has been used to study rich and nontrivial physics such as superradiance and quantum phase transitions \cite{Emary_and_Brandes_2003_QPT_Dicke_PRL,Emary_and_Brandes_2003_QPT_Dicke_PRE}. Such physics have been found to occur in the ultrastrong coupling (USC) regime \cite{Garbe_et_al_2017_SPT_in_USC,Bin_et_al_2019_Collective_radiance_in_USC,Frisk_Kockum_2019_Ultrastrong_coupling} where the field-atom coupling coefficient is comparable to the atomic transition frequency. In this regime, the rotating wave approximation is invalid, rendering the analysis of the system much more difficult. Hence, approximate techniques such as the Holstein-Primakoff transformation are often used, which is valid only in specific settings.

Nevertheless, the USC regime gives rise to intriguing physics that must be explored further. For instance, it may enable fast two-qubit gate operations for quantum computing applications \cite{Stassi_Cirio_&_Nori_2020_Scalable_QC_in_USC_regime}, and it has been shown that the multimode fields must be taken into account in order for the system to be causal \cite{Munoz_superluminal_2018}. Furthermore, from rigorous derivation in circuit quantum electrodynamics (QED), the extended Dicke model has been shown to include a direct qubit-qubit interaction term which becomes non-negligible in the USC regime \cite{Jaako_et_al_2016_USC_phenomena_beyond_the_Dicke_model}. Such novel phenomena unique to this regime greatly motivate the need to develop numerical techniques to efficiently simulate multi-atom, multimode systems.

For the study of single atom interacting with multiple field modes, the chain mapping technique has been extremely useful for tensor network analysis of the spin-boson model \cite{Bulla_et_al_2005_NRG,Bulla_et_al_2008_NRG,Prior_et_al_2010_Strong_system-environment_interactions,Chin_et_al_2010_exact_transformation_to_chain} and multimode quantum Rabi model \cite{Munoz_superluminal_2018,Ryu_Na_Chew_2023_MPS_and_NMD}. Since these models have the so-called star coupling structure \cite{Bulla_et_al_2005_NRG}, they are transformed to an equivalent Hamiltonian with a linear chain coupling structure with nearest-neighbor interactions. Once transformed, numerical algorithms such as matrix product states (MPS) \cite{Vidal_2003_MPS,Vidal_2004_MPS} or density matrix renormalization group \cite{White_1992_DMRG} can be applied efficiently.

Although it is highly effective, the chain mapping technique is limited to systems with single two-level atom or spin-1/2 system. Naturally, various efforts have been made to extend the technique to more general systems such as the two-bath spin-boson model \cite{Guo_et_al_2012_Two-bath_SBM,Dunnett_and_Chin_2021_Two-bath_SBM}. However, the generalization to a multi-atom, multimode system has remained challenging. Strathearn \textit{et al}.\ \cite{Strathearn_et_al_2018_TEMPO} developed an extension to a spin-boson model with two spins by projecting the system onto a subspace and mapping it to a single spin-boson model. Most notably, transformation of the multi-spin-boson model to a chain-like structure has been achieved by applying the block Lanczos algorithm \cite{Noachtar_et_al_2022_Giant_atoms}.

In this paper, we present a novel generalization of the chain mapping technique that works for few-atom, multimode systems. Our method is straightforward to implement and numerically stable as opposed to methods involving Lanczos algorithms. It also completely specifies all coupling coefficients (field-atom and field-field) after the transformation for arbitrary configurations. Our method utilizes coupling matrix transformations to achieve this, and it leads to equivalent Hamiltonians that are more compatible with tensor network algorithms. We demonstrate this through various numerical simulations in the USC regime.

\begin{figure*}
	\includegraphics[width=1\linewidth]{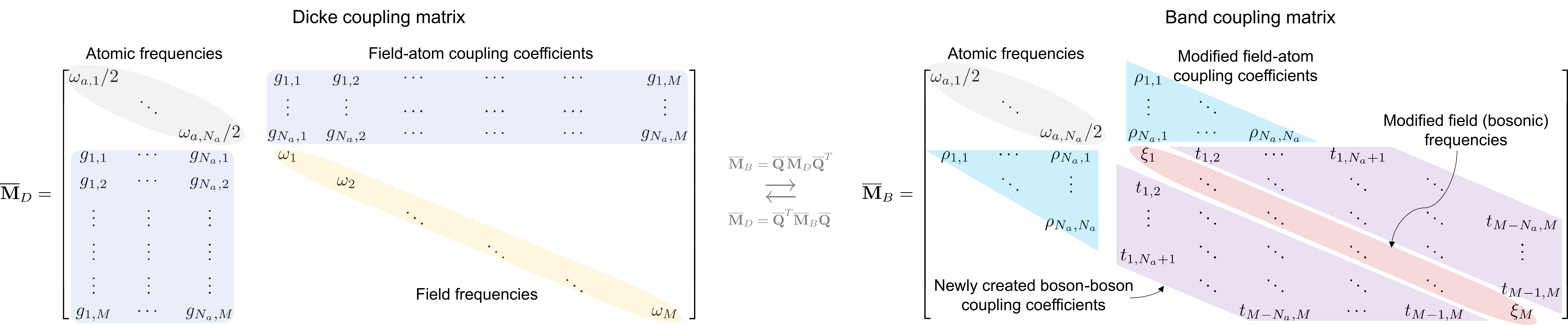}
	\caption{The coupling matrices for the Dicke (left) and band (right) Hamiltonians. The matrix entries correspond to the coefficients in (\ref{eq:Dicke_Ham}) and (\ref{eq:band_Ham}).}
	\label{fig:coup_mats}
\end{figure*}

The rest of this paper is organized as follows. In Sec.~\ref{sec:gen_chain_mapping}, we present our formulation that is based on coupling matrix transformations and discuss the applicability and limitations of our approach. After that, in Sec.~\ref{sec:num_validation}, we numerically validate our proposed approach. A numerical example of two entangled atoms ultrastrongly coupled to multimode fields is discussed in Sec.~\ref{sec:num_ex}\@. And finally, summary and directions for future work are given in Sec.~\ref{sec:conc}.

\section{Generalized chain mapping}\label{sec:gen_chain_mapping}
\subsection{Formulation}
We are primarily concerned with QED applications in the USC regime where the rotating wave approximation is invalid. Therefore, we use the multimode Dicke Hamiltonian to represent a system with $N_a$ atoms and $M$ electromagnetic modes, namely
\begin{equation}\label{eq:Dicke_Ham}
\hat{H}_D=\hbar\sum_{j=1}^{N_a}\frac{\omega_{a,j}}{2}\hat\sigma^z_j+\hbar\sum_{k=1}^{M}\Big[\omega_k\hat{a}_k^\dagger\hat{a}_k-i\sum_{j=1}^{N_a}g_{j,k}\hat\sigma^x_j(\hat{a}_k-\hat{a}_k^\dagger)\Big],
\end{equation}
where $\hbar$ is the reduced Planck constant; $\hbar\omega_{a,j}$ and $\hat\sigma^l_j$ are the energy gap and the Pauli operator, respectively, of the $j$-th atom where $l=x,y,$ or $z$; $\omega_k$ is the electromagnetic mode frequency, and $\hat{a}_k$ ($\hat{a}_k^\dagger$) is the photon annihilation (creation) operator, all for mode-$k$. The coupling coefficient between the $j$-th atom and mode-$k$ is $g_{j,k}$, which is based on the electric dipole interaction \cite{Ryu_Na_Chew_2023_MPS_and_NMD}.

The coupling structure of the multimode Dicke Hamiltonian can be represented by a coupling matrix of size $(N_a+M)\times(N_a+M)$ partitioned as
\begin{equation}
\overline{\bf M}_D=
\begin{bmatrix}
\overline{\boldsymbol{\omega}}_a & \overline{\bf g} \\
\overline{\bf g}^T & \overline{\boldsymbol{\omega}}_f
\end{bmatrix},
\end{equation}
where $\overline{\boldsymbol{\omega}}_a$ and $\overline{\boldsymbol{\omega}}_f$ are diagonal matrices of atomic and field frequencies, and $\overline{\bf g}$ is an $N_a\times M$ dense matrix representing the field-atom coupling coefficients. The Dicke coupling matrix is a real-valued, symmetric matrix that is visualized on the left in Fig.~\ref{fig:coup_mats}.

The far off-diagonal coupling elements of $\overline{\bf M}_D$ such as $g_{1,M}$ (shown in Fig.~\ref{fig:coup_mats}) are what makes the tensor network simulation of (\ref{eq:Dicke_Ham}) inefficient. In MPS simulations, this type of interaction terms are referred to as long-range interactions \cite{Haegeman_et_al_2016_TDVP}, and they require implementing a great number of SWAP gates.\footnote{A SWAP gate is a quantum gate that exchanges the states of two qubits. More generally, for MPS simulations, a SWAP gate exchanges the states of two identical quantum systems \cite{Stoudenmire_and_White_2010_MPS_zip-up}.} To avoid this inefficiency, we annihilate these coupling elements by applying a series of Householder transformations {\color{red}\cite{Householder_1958_Unitary_triangularization,Heath_Sci_Comp}} and orthogonally transform the coupling matrix into a band matrix as
\begin{equation}\label{eq:CMT}
\overline{\bf M}_B=\underbrace{\overline{\bf Q}_{M-N_a-1}\dots\overline{\bf Q}_2\overline{\bf Q}_1}_{=\overline{\bf Q}}\overline{\bf M}_D\overline{\bf Q}_1^T\overline{\bf Q}_2^T\dots\overline{\bf Q}_{M-N_a-1}^T
\end{equation}
or simply $\overline{\bf M}_B=\overline{\bf Q}\,\overline{\bf M}_D\overline{\bf Q}^T$.

The Householder matrix is constructed as
\begin{equation}
\overline{\bf Q}_i=\overline{\bf I}-\frac{2{\bf v}_i{\bf v}_i^T}{{\bf v}_i^T{\bf v}_i}
\end{equation}
where $\overline{\bf I}$ is an $(N_a+M)\times(N_a+M)$ identity matrix, and the vector ${\bf v}_i$ is built from the $i$-th column of $\overline{\bf M}_D^{(i-1)}$ as
\begin{equation}\label{eq:Householder_vector}
{\bf v}_i=
\begin{bmatrix}
{\bf 0}_i \\ {\bf m}_i
\end{bmatrix}
-\alpha{\bf e}_{N_a+i}.
\end{equation}
In the above, $\overline{\bf M}_D^{(i-1)}=\overline{\bf Q}_{i-1}\dots\overline{\bf Q}_1\overline{\bf M}_D\overline{\bf Q}_1^T\dots\overline{\bf Q}_{i-1}^T$; ${\bf 0}_i$ is a length-$(N_a+i-1)$ vector of zeros; ${\bf m}_i$ is a length-$(M-i+1)$ vector whose entries are equal to everything below $(N_a+i-1)$-th entry of the $i$-th column of $\overline{\bf M}_D^{(i-1)}$; $\alpha=-\operatorname{sgn}(m_{N_a+i,i}^{(i-1)})\lVert{\bf m}_i\rVert_2$ with sgn being the sign function that is defined to be one at the origin [$\operatorname{sgn}(0)=1$] and $\lVert\,\cdot\,\rVert_2$ being the 2-norm; and ${\bf e}_{N_a+i}$ is a standard unit vector that is equal to one in the $(N_a+i)$-th entry with zeros in all other entries. Equation (\ref{eq:Householder_vector}) and $\overline{\bf M}_D^{(i-1)}$ are illustrated in Fig.~\ref{eq:Householder_vector} for visual clarity.

\begin{figure}
	\includegraphics[width=.75\linewidth]{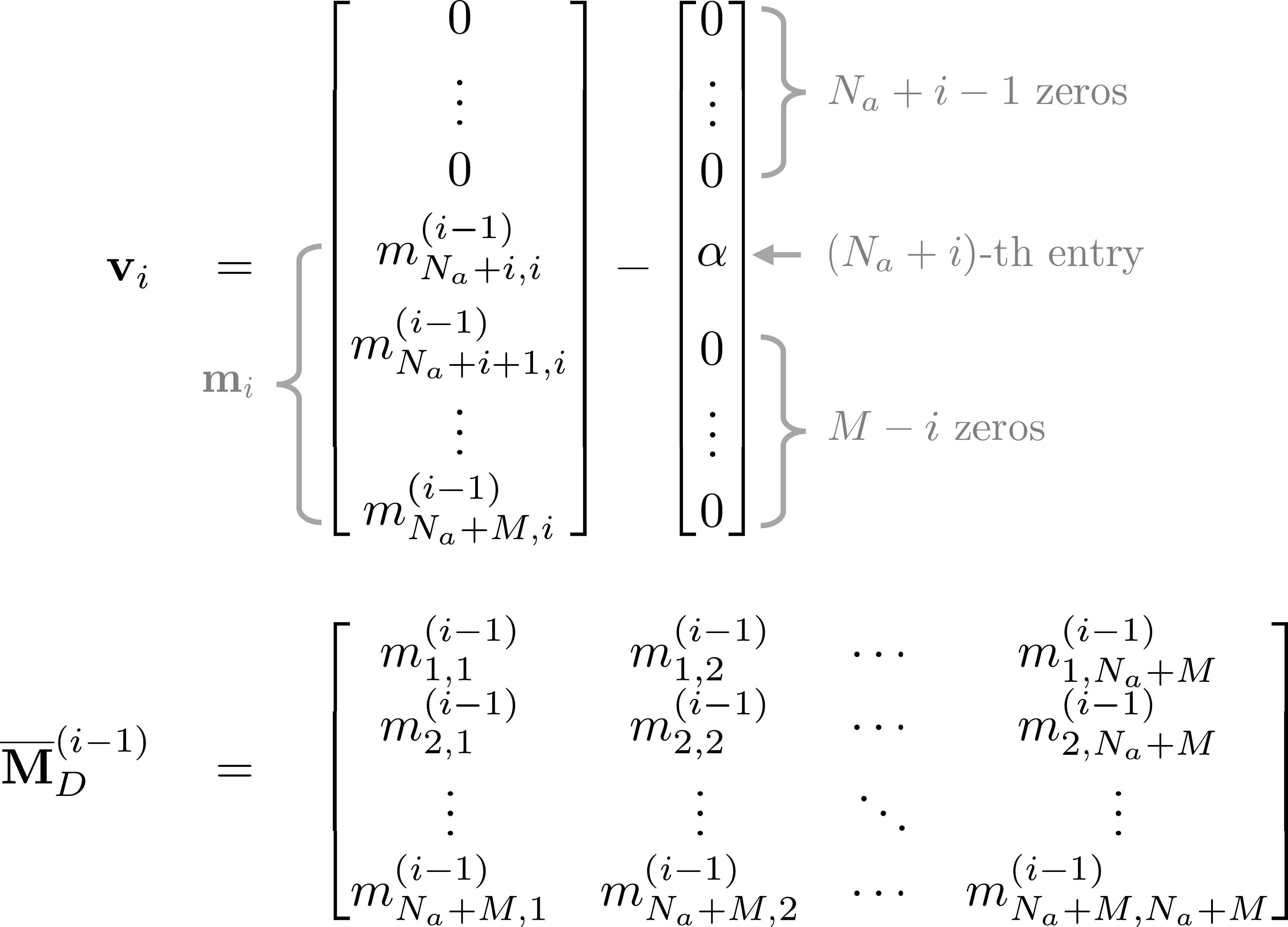}
	\caption{The vector ${\bf v}_i$ (\ref{eq:Householder_vector}) is built from a part of the $i$-th column of the intermediate matrix $\overline{\bf M}_D^{(i-1)}$.}
	\label{fig:Householder_vector}
\end{figure}

In (\ref{eq:CMT}), each $\overline{\bf Q}_i$ implements a Householder transformation that annihilates everything below the $(N_a+i)$-th entry of the $i$-th column of the target matrix $\overline{\bf M}_D^{(i-1)}$. The final result is a symmetric band matrix of the same size as $\overline{\bf M}_D$ with bandwidth $N_a$ that can be expressed in the block matrix form as
\begin{equation}
\overline{\bf M}_B=
\begin{bmatrix}
\overline{\boldsymbol{\omega}}_a & \overline{\boldsymbol{\rho}} \\
\overline{\boldsymbol{\rho}}^T & \overline{\boldsymbol{\xi}}
\end{bmatrix},
\end{equation}
where $\overline{\boldsymbol{\rho}}$ is a lower-triangular matrix of size $N_a\times M$ representing the modified field-atom coupling coefficients, and $\overline{\boldsymbol{\xi}}$ is a band matrix of size $M\times M$ with diagonal elements $[\overline{\boldsymbol{\xi}}]_{ii}=\xi_i$ representing the transformed bosonic frequencies and off-diagonal elements $[\overline{\boldsymbol{\xi}}]_{ij}=t_{ij}$ for $i\ne j$ and $|i-j|\le N_a$ representing the newly created boson-boson coupling coefficients.
The resulting band matrix is visualized on the right in Fig.~\ref{fig:coup_mats}.

Inspired by the chain mapping technique \cite{Bulla_et_al_2005_NRG,Bulla_et_al_2008_NRG,Prior_et_al_2010_Strong_system-environment_interactions,Chin_et_al_2010_exact_transformation_to_chain}, the Hamiltonian whose coupling structure is represented by the band coupling matrix $\overline{\bf M}_B$ can be written in terms of the entries of $\overline{\bf M}_B$ as
\begin{equation}\label{eq:band_Ham}
\begin{aligned}
\hat{H}_B&=\hbar\sum_{j=1}^{N_a}\bigg[\frac{\omega_{a,j}}{2}\hat\sigma^z_j-i\sum_{k\le j}\rho_{j,k}\hat\sigma^x_j(\hat{b}_k-\hat{b}_k^\dagger)\bigg]\\
&\quad+\hbar\sum_{k=1}^{M}\Big[\xi_k\hat{b}_k^\dagger\hat{b}_k+\sum_{j=1}^{N_a}t_{k,k+j}(\hat{b}_k^\dagger\hat{b}_{k+j}+\hat{b}_{k+j}^\dagger\hat{b}_k)\Big]
\end{aligned}
\end{equation}
with $t_{k,k+j}=0$ for $k+j>M$. It is remarkable that (\ref{eq:band_Ham}) compared to (\ref{eq:Dicke_Ham}) lacks the far off-diagonal couplings as a result of the coupling matrix transformation. This is explicitly shown by the summation indices for the interaction terms that are limited by the atomic index $j$ (first row, second summation) and the number of atoms $N_a$ (second row, second summation). This absence of far off-diagonal couplings is what makes (\ref{eq:band_Ham}) much more compatible with tensor network algorithms.

The orthogonal matrix $\overline{\bf Q}$ in (\ref{eq:CMT}) that implements the transformation is of the form
\begin{equation}\label{eq:Q_def}
\overline{\bf Q}=
\begin{bmatrix}
\overline{\bf I}_{N_a} & \overline{\bf 0} \\
\overline{\bf 0} & \overline{\bf U}
\end{bmatrix},
\end{equation}
where $\overline{\bf I}_{N_a}$ is an $N_a\times N_a$ identity matrix, and $\overline{\bf U}$ is an $M\times M$ orthogonal matrix. From this form of $\overline{\bf Q}$, it is evident that the transformation only applies to the bosons and not the atoms. This is why the atomic frequencies in $\overline{\bf M}_B$ are left unchanged and are equal to those in $\overline{\bf M}_D$. The photonic operator $\hat{a}_k$ and the chain bosonic operator $\hat{b}_j$ are related as $\hat{b}_j=\sum_{k=1}^{M}U_{jk}\hat{a}_k$ where $U_{jk}=[\overline{\bf U}]_{jk}$ is the block matrix from (\ref{eq:Q_def}). In other words, a particular way of clustering the photons gives rise to the chain bosonic modes. This is precisely the same as what is done in the chain mapping technique \cite{Bulla_et_al_2005_NRG,Bulla_et_al_2008_NRG,Prior_et_al_2010_Strong_system-environment_interactions,Chin_et_al_2010_exact_transformation_to_chain}.

There is a good reason that we do not trim the off-diagonal elements of $\overline{\bf M}_D$ all the way to the tridiagonal form. If it were tridiagonalized, then we would lose the identity block matrix $\overline{\bf I}_{N_a}$ in the upper left corner of (\ref{eq:Q_def}), and $\overline{\bf Q}$ will end up being a full matrix, which would mix the two-level and bosonic operators in the process of transformation. The resulting Hamiltonian will not be well-defined. To avoid this, we make sure that the transformation only applies to the bosonic operators as shown in (\ref{eq:Q_def}).



\subsection{Complexity, applicability, and limitations}
Since our scheme is based on Householder transformations which costs $O(N^3)$ where $N=N_a+M$ is the number of rows and columns of the coupling matrices, it is numerically stable. This is in contrast to the chain mapping technique that is unstable due to the Lanczos algorithm [$O(MN^2)$ {assuming $M>N_a$] on which it is based. Hence, chain mapping needs stabilization using methods such as the modified Gram-Schmidt orthogonalization \cite{Ryu_Na_Chew_2023_MPS_and_NMD} which costs $O(N^3)$ also.

The generalized chain mapping technique works the best when there are just few atoms interacting with multiple modes ($N_a<M$). If $N_a\ge M$, then $N_a-M+1$ atoms remain coupled to all the modes even after the coupling matrix transformation, making the resulting system incompatible with tensor network algorithms.

We have investigated time-domain MPS simulations with the band Hamiltonian (\ref{eq:band_Ham}) using the time-evolving block decimation (TEBD) algorithm \cite{Vidal_2004_MPS,Paeckel_et_al2019_TE_methods_for_MPS} and the time-dependent variational principle \cite{Haegeman_et_al_2011_TDVP,Haegeman_et_al_2016_TDVP}. Both algorithms require forming an $(N_a+1)$-site operator due to the near-neighbor coupling structure of the band Hamiltonian (\ref{eq:band_Ham}). For example, in TEBD, this operator is then turned into a matrix product operator (MPO), whose bond dimensions scale exponentially in $N_a$. If we let $N_f$ be the number of Fock states considered for each bosonic mode in the simulation, then the maximum bond dimension of the MPO would be $N_f^{N_a}$ or $N_f^{N_a+1}$ when $N_a$ is even or odd, respectively. An example of $N_a=3$ case is illustrated in Fig.~\ref{fig:MPO_bond_dim}.
\begin{figure}
	\includegraphics[width=.9\linewidth]{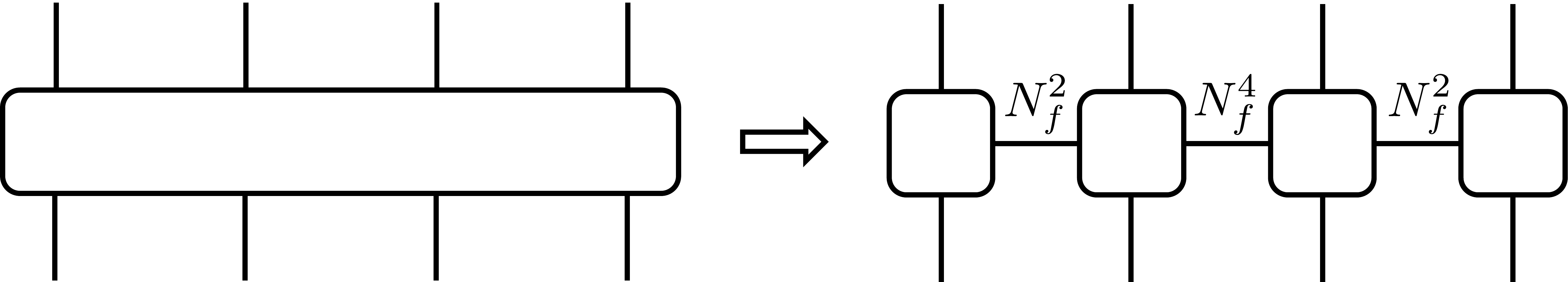}
	\caption{Left: 4-site operator for TEBD in $N_a=3$ case. Right: Decomposed MPO of the 4-site operator. All vertical indices (on top and bottom of tensors) are bosonic indices of dimension $N_f$. As a result of the decomposition, bonds are formed with the highest dimensional bond being at the center of the MPO.}
	\label{fig:MPO_bond_dim}
\end{figure}
The 4-site operator in this example is formed from a part of the band Hamiltonian (\ref{eq:band_Ham}) as $\exp\{-i\hat{h}_k\Delta t/\hbar\}$ where $\hat{h}_k=\;$ $\xi_k\hat{b}_k^\dagger\hat{b}_k+\sum_{j=1}^{3}t_{k,k+j}(\hat{b}_k^\dagger\hat{b}_{k+j}+\hat{b}_{k+j}^\dagger\hat{b}_k)$ for some $k$.

At the onset of the USC regime, we found that at least $N_f=8$ is required for the simulation times we explore in Sec.~\ref{sec:num_ex}\@. In this case, the largest bond dimension of the MPO would be $N_f^4=8^4=4096$, which makes 3-atom MPS simulation computationally impracticable at this coupling strength. For weaker coupling strengths where $N_f$ could be lower without sacrificing the accuracy, it is possible to simulate up to 3 or 4 atoms with this approach. For these reasons, we simulated two atoms in the USC regime as demonstrated in Sec.~\ref{sec:num_ex}.

\section{Numerical validation}\label{sec:num_validation}
\subsection{Equivalence to the chain mapping}
What is reassuring is that when $N_a=1$, our coupling matrix transformation technique reduces to the chain mapping technique and implements the exact same transformation. This is numerically demonstrated\footnote{Since the Householder transformation is by and large a numerical technique, it is difficult (or impossible) to mathematically prove the equivalence between our coupling matrix transformation and the chain mapping technique. This is why we numerically demonstrate their equivalence.} in Fig.~\ref{fig:plot_orig_vs_new_transformation}. Here, we consider a single atom placed at the center of a 1D lattice with periodic boundary conditions, and we consider the lowest fifty electromagnetic eigenmodes of this system. The parameters for this test are selected to be $\omega_k=k\omega_a$ with integer $k\in[1,50]$ and $g_k=g\sqrt{\omega_k}$ with $g=1$. Since there is only one atom in this case, the atomic index is fixed at $j=1$ and is omitted here.  It is observed in Fig.~\ref{fig:plot_orig_vs_new_transformation} that the coefficients agree perfectly, so it is clear that the coupling matrix transformation reduces to the chain mapping technique in the single atom case.

\begin{figure}
	\includegraphics[width=1\linewidth]{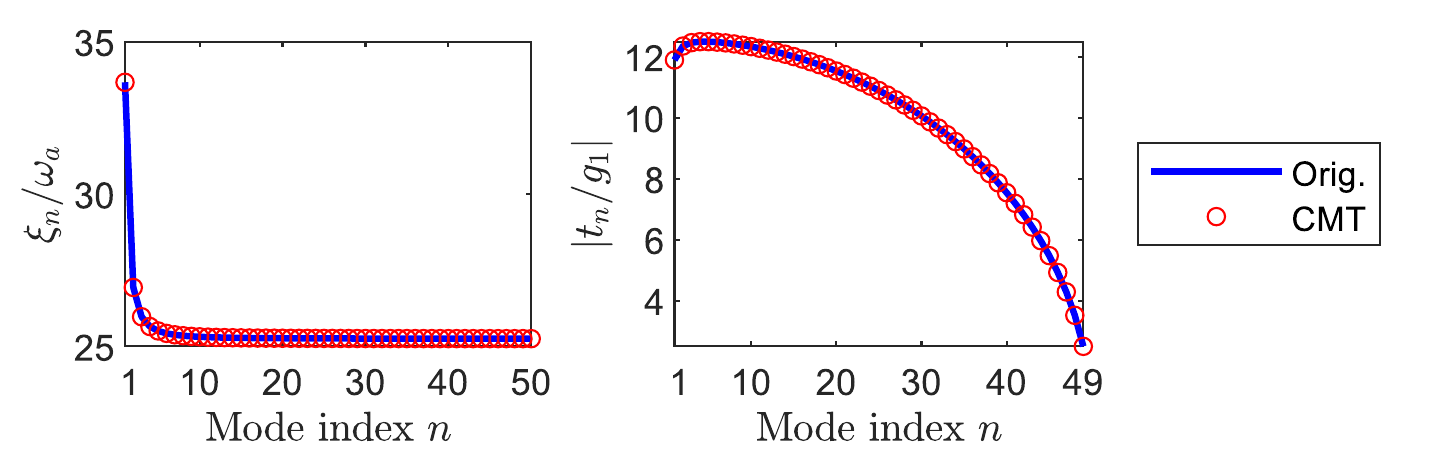}
	\caption{Comparison of the original chain mapping technique from \cite{Bulla_et_al_2005_NRG} and our coupling matrix transformation (CMT) approach presented in this paper. They are in good agreement. $\xi_n$ represents the transformed bosonic frequencies, and $t_n$ represents the chain boson-boson coupling coefficient.}
	\label{fig:plot_orig_vs_new_transformation}
\end{figure}

\subsection{Equivalence of the Hamiltonians}
To show that the multimode Dicke Hamiltonian (\ref{eq:Dicke_Ham}) is equivalent to the band Hamiltonian (\ref{eq:band_Ham}), we perform a simple numerical time-domain simulation for both systems for 3-atom, 5-mode case in the USC regime.\footnote{This is realized by setting one of the coupling coefficients as $g_{1,1}/\omega_1=0.25$, i.e., the normalized coupling coefficient between the first atom and fundamental field mode of the cavity is 0.25. The coupling coefficients for the other modes and other atoms are determined by the electric dipole interaction which depends on the field profile, atoms' positions, and their dipole moments. Since we assume identical atoms here, their dipole moments are equal. The expression for the electric dipole coupling coefficient is given in \cite[Eq.~(7)]{Ryu_Na_Chew_2023_MPS_and_NMD}.} This is small enough that the computational cost for simulating either system is very low (and tensor network algorithm is not needed here).

The three atoms are assumed to be identical and placed in a 1D cavity with perfect electric conductor (PEC) walls. The cavity occupies $x\in[-L/2,L/2]$ where $L$ is the length of the cavity, and the atoms are placed at $x=0$, $L/4$, and $-3L/8$. This setting is depicted in Fig.~\ref{fig:three_atoms_in_cav}. The plot of the time-domain simulation result is shown in Fig.~\ref{fig:plot_Dicke_vs_band}. We numerically solve the quantum state equation (also known as Schr\"odinger equation) for Hamiltonians (\ref{eq:Dicke_Ham}) and (\ref{eq:band_Ham}) to obtain the time-evolved state $\ket{\psi(t)}$ and compute the excited-state atomic population $\braket{\sigma^+_j\sigma^-_j}=\braket{\psi(t)|\hat\sigma^+_j\hat\sigma^-_j|\psi(t)}$ for each atom. The initial state is given by $\ket{\psi_0}=\ket{e,e,e,0,\dots,0}$, i.e., three excited atoms in vacuum. Excellent agreement is observed in Fig.~\ref{fig:plot_Dicke_vs_band}, and we conclude that the Dicke (\ref{eq:Dicke_Ham}) and band (\ref{eq:band_Ham}) Hamiltonians represent an equivalent physical system.

\begin{figure}
\includegraphics[width=.5\linewidth]{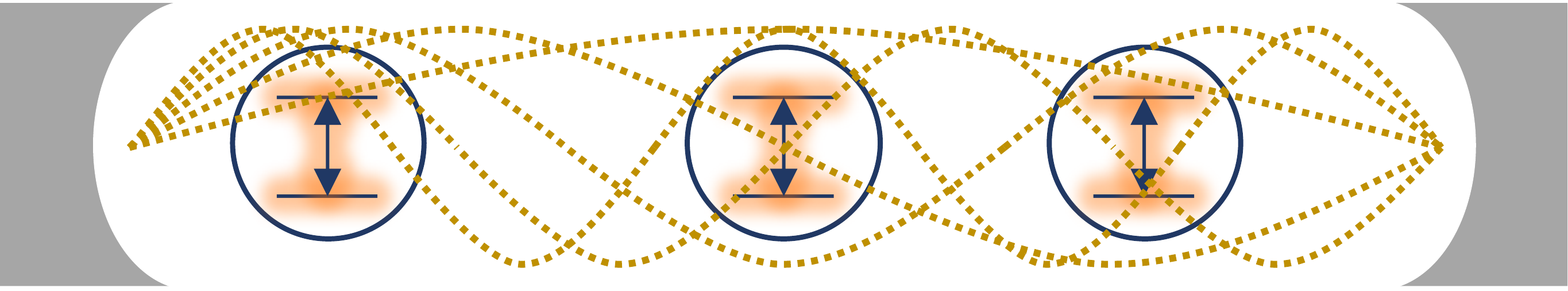}
\caption{Illustration of three identical atoms placed in a 1D PEC cavity interacting with the first five modes.}
\label{fig:three_atoms_in_cav}
\end{figure}

\begin{figure}
\includegraphics[width=1\linewidth]{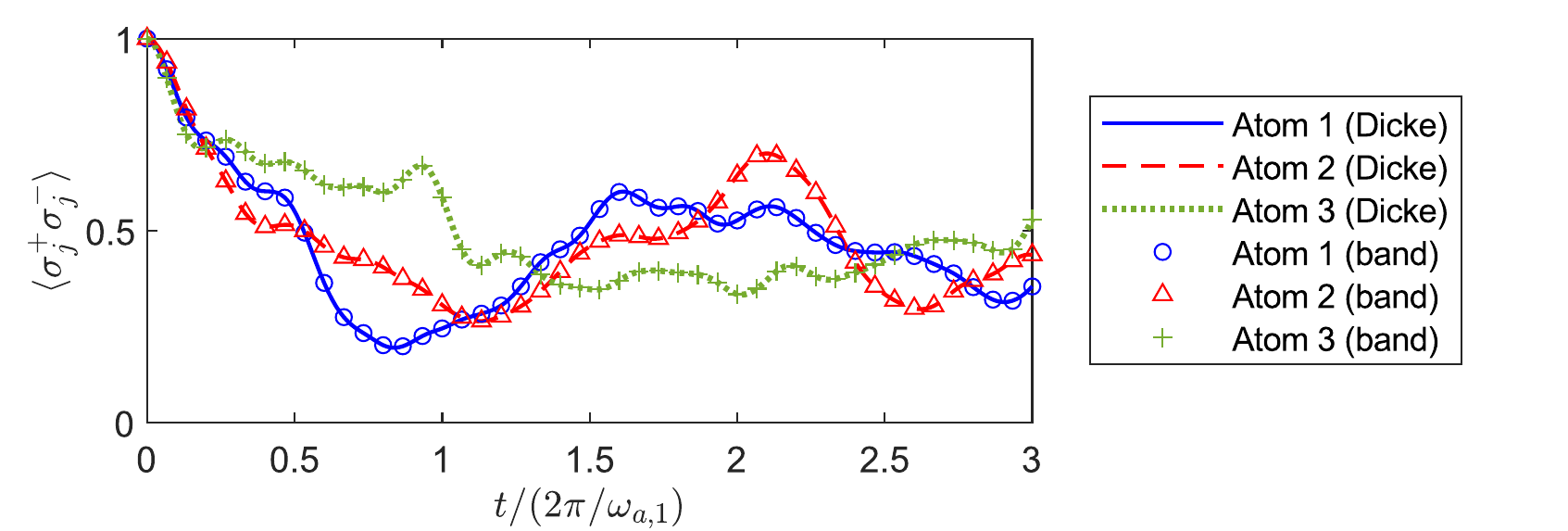}
\caption{3-atom, 5-mode simulation of the Dicke (\ref{eq:Dicke_Ham}) and band (\ref{eq:band_Ham}) Hamiltonians with normalized coupling strength 0.25 between Atom 1 and the fundamental mode of the cavity. Here, we are numerically demonstrating the equivalence between the two Hamiltonians.}
\label{fig:plot_Dicke_vs_band}
\end{figure}

\begin{figure*}
\includegraphics[width=1\linewidth]{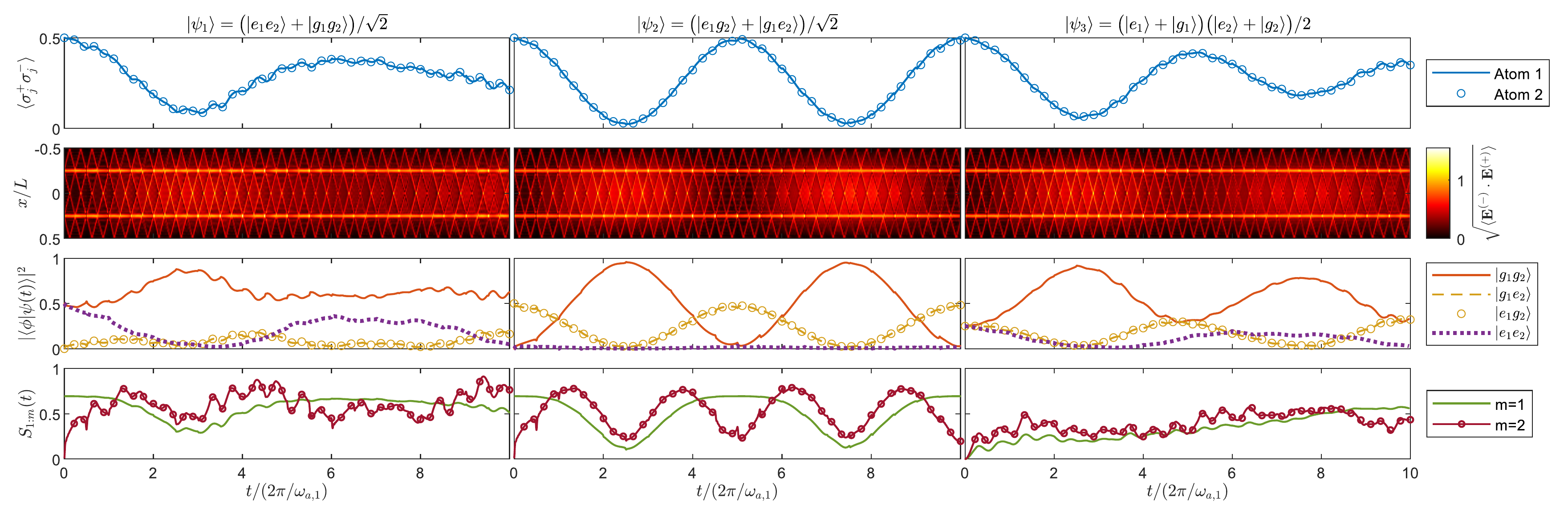}
\caption{MPS simulation results of the two-atom, thirty-mode Dicke model using the band Hamiltonian ({\ref{eq:band_Ham}}). First, the atomic population in the excited state is calculated as $\braket{\sigma_j^+\sigma_j^-}=\braket{\psi(t)|\hat\sigma_j^+\hat\sigma_j^-|\psi(t)}$ and plotted in the top row. This is equal for both atoms because they are placed symmetrically within the cavity. Second, the first-order field correlation function is computed as $\braket{{\bf E}^{(-)}\cdot{\bf E}^{(+)}}=\braket{\psi(t)|\hat{\bf E}^{(-)}({\bf r})\cdot\hat{\bf E}^{(+)}({\bf r})|\psi(t)}$ whose calculation is detailed in \cite{Ryu_Na_Chew_2023_MPS_and_NMD}. Third, we compute the components of four possible two-atomic states in $\ket{\psi(t)}$ as $|\braket{\phi|\psi(t)}|^2$ where $\ket{\phi}=\ket{g_1g_2}$, $\ket{g_1e_2}$, $\ket{e_1g_2}$, and $\ket{e_1e_2}$. Finally, the von Neumann entanglement entropy of the MPS $S_{1:m}(t)$ is calculated for two different bipartitions of the MPS: one between the first two sites and the other between the second and third sites.}
\label{fig:ent_sim_results}
\end{figure*}

\section{Numerical example: Two entangled atoms in a cavity}\label{sec:num_ex}
In the USC regime, single electromagnetic mode approximation is likely to fail due to the possibility of superluminal signaling \cite{Munoz_superluminal_2018}, so multiple modes must be considered. It was found that several tens of modes are enough to accurately characterize the propagation effects in this coupling regime \cite{Munoz_superluminal_2018}. When so many modes need to be incorporated into the model, it is highly inefficient (or impossible) to numerically simulate the system using the multimode Dicke Hamiltonian (\ref{eq:Dicke_Ham}). Since we have seen their equivalence, we use the band Hamiltonian (\ref{eq:band_Ham}) for tensor network simulations in the remainder of this paper.

\subsection{Simulation setting}
We restrict our MPS simulations to two identical atoms in 1D PEC cavity in the presence of thirty electromagnetic modes ($M=30$). In particular, we are interested in how entangled atoms interact with multiple field modes in the USC regime. This is important since obtaining entangled qubits is an essential step in all quantum algorithms. Moreover, it has been shown in circuit QED that ultrastrong interactions can be leveraged to ``harvest'' entangled atoms \cite{Armata_et_al_2017_Harvesting_multiqubit_ent}. Our simulations will show how quantized multimode fields impact the time evolution of entangled atoms in different configurations in the USC regime.

The simulation setting is similar to the one depicted in Fig.~\ref{fig:three_atoms_in_cav} but only with two atoms now placed at $x=\pm L/4$. We assume that the atoms are resonant with the fundamental mode of the cavity such that the mode frequencies are $\omega_k=(2k-1)\omega_a$ with integer $k\in[1,M]$, where the atomic frequency for both atoms is $\omega_a$. These mode frequencies are a consequence of the homogeneous 1D PEC cavity. Other than the fact that it eliminates the possibility of superluminal signaling \cite{Munoz_superluminal_2018}, considering 30 modes is also adequate because in this type of setting the normalized coupling coefficient ($g_{j,k}/\omega_k$) decays rapidly with mode index $k$ \cite{Ryu_Na_Chew_2023_MPS_and_NMD}. MPS is used to efficiently represent the time-evolving quantum state $\ket{\psi(t)}$, and the time evolution operator $e^{-i\hat{H}_B\Delta t/\hbar}$ is approximately constructed as an MPO using TEBD.

We consider three initial states:
\begin{subequations}
\begin{align}
\ket{\psi_1}&=\big(\ket{e_1e_2}+\ket{g_1g_2}\big)/\sqrt{2},\label{eq:psi_1}\\
\ket{\psi_2}&=\big(\ket{e_1g_2}+\ket{g_1e_2}\big)/\sqrt{2},\label{eq:psi_2}\\
\ket{\psi_3}&=\big(\ket{e_1}+\ket{g_1}\big)\big(\ket{e_2}+\ket{g_2}\big)/2\label{eq:psi_3}
\end{align}
\end{subequations}
with vacuum (no photons) in all three cases. The subscripts in the above distinguish the two atoms. The first two states are maximally entangled with different configurations, while the last is separable (non-entangled). We reveal the differences in their time evolution characteristics for these three initial states.

\subsection{MPS simulation results}
We deal with the onset of the USC regime where $\max_{j,k}|g_{j,k}/\omega_k|=0.1$. The simulation results are shown in Fig.~\ref{fig:ent_sim_results}. In particular, the von Neumann entanglement entropies are plotted in the last row to quantify the degree of entanglement for two different bipartitions of the time-evolving MPS. The entropies are expressed as
\begin{subequations}
\begin{align}
S_1(t)&=-\operatorname{tr}[\hat\rho_1(t)\ln\hat\rho_1(t)],\\
S_{1:2}(t)&=-\operatorname{tr}[\hat\rho_{1:2}(t)\ln\hat\rho_{1:2}(t)],
\end{align}
\end{subequations}
where the reduced density operators in the above are obtained by taking the partial trace of the total density operator, $\rho(t)=\ket{\psi(t)}\bra{\psi(t)}$, as $\hat\rho_{1:m}(t)=\operatorname{tr}_{m+1:N}[\hat\rho(t)]$, where $m$ indexes the physical sites of MPS, and the bipartition is taken between sites $m$ and $m+1$. When $m=1$, we simply denote $\hat\rho_{1:1}(t)=\hat\rho_1(t)$. The density operator and its bipartitions are further explained in Fig.~\ref{fig:mps_dens_op_bipartition}. With MPS, these entropies can be simply computed by taking the singular values at the zero-site center located right along the bipartition \cite{Haegeman_et_al_2016_TDVP}.

\begin{figure}
\includegraphics[width=.55\linewidth]{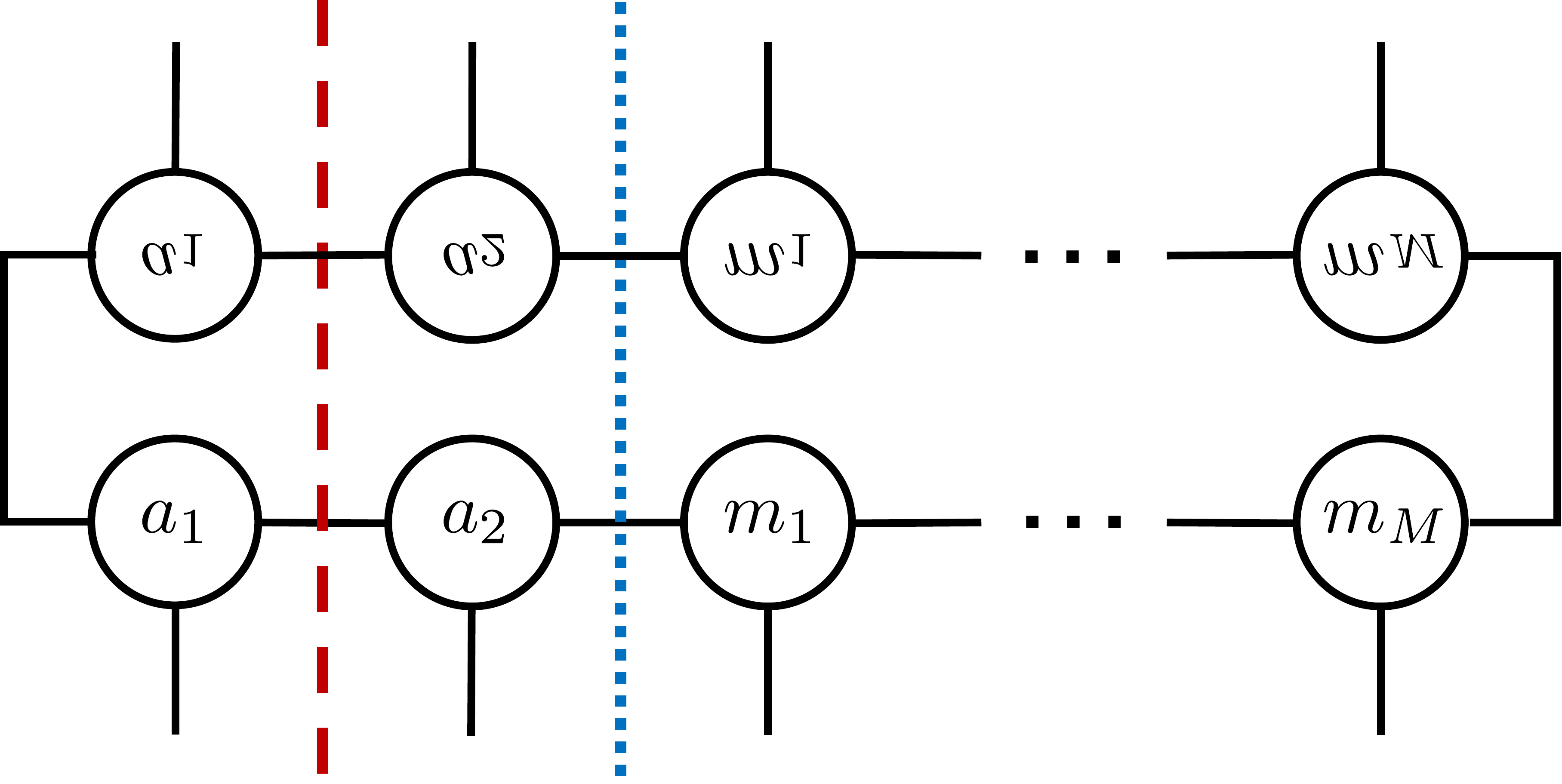}
\caption{Tensor network diagram of the total density operator formed by taking the outer product of the MPS. The first two sites represent the atoms, and the remaining sites represent the electromagnetic modes. The bipartition is taken between sites one and two (red dashed line) to calculate $S_1(t)$, and sites two and three (blue dotted line) to calculate $S_{1:2}(t)$. After the bipartition, the right side of the density operator is traced out by taking the partial trace.}
\label{fig:mps_dens_op_bipartition}
\end{figure}

The simulations in Fig.~\ref{fig:ent_sim_results} take place at the lowest end of the USC regime where both the weak and USC effects take place. The weak coupling effect is shown in the field correlation plot where a ``glow'' in the cavity is observed. This glow represents the fundamental mode of the cavity to which the atoms couple dominantly. The propagation effects characterized by the traveling wavefront (traveling at the speed of light) is also visible due to the USC between the atoms and field modes.

What is notable about the simulation results in Fig.~\ref{fig:ent_sim_results} is that although both initial states (\ref{eq:psi_1}) and (\ref{eq:psi_2}) are maximally entangled states, they exhibit very different behaviors in the presence of multiple electromagnetic modes. The initial state (\ref{eq:psi_2}) displays a highly periodic behavior resembling Rabi oscillations. It can be seen that this initial state is almost fully revived when $t/(2\pi/\omega_{a,1})=5$, meaning that both atoms nearly go back to the maximally entangled initial state. This does not happen for initial states (\ref{eq:psi_1}) and (\ref{eq:psi_3}).

Regarding the von Neumann entanglement entropy, the maximum possible value of $S_1(t)$ is $\ln2\approx0.693$ since the first site of the MPS is occupied by a two-level atom. When $S_1(t)$ goes back close to the maximum value at times $t>0$ for the entangled initial states (\ref{eq:psi_1}) and (\ref{eq:psi_2}), either the atoms are back to the maximally entangled initial state (\ref{eq:psi_2}), or the first atom is entangled with both the second atom and the field modes in the case of (\ref{eq:psi_1}). For the separable initial state (\ref{eq:psi_3}), we observe the entropies starting out at zero and slowly increasing over time. For long enough simulations, these values will saturate to a level that depends on the coupling strength.

\section{Summary and future work}\label{sec:conc}
We have presented a novel generalization of the chain mapping technique based on coupling matrix transformations that works accurately for few-atom, multimode systems. Our technique is very useful for tensor network simulations of the multimode Dicke model and multi-spin-boson model because it can take the coupling structures of these models and alter them into a linear chain form with near-neighbor interactions, which is highly compatible with MPS\@. The coupling matrix transformations are numerically stable, and this technique reduces to the chain mapping technique in the single atom case. We have demonstrated the equivalence between the Dicke (\ref{eq:Dicke_Ham}) and band (\ref{eq:band_Ham}) Hamiltonians and applied the band Hamiltonian for MPS simulations of two entangled atoms with thirty electromagnetic modes. Our future work involves extending this technique for realistic 3D models such as flux qubits ultrastrongly coupled to coplanar waveguide resonators \cite{Niemczyk_et_al_2010_Circuit_QED_in_the_USC_regime,Wang_et_al_2020_Bloch-Siegert_shift_in_US_coupled_circuit_QED}.

\begin{acknowledgments}
This work was supported by the National Science Foundation Grant No.~2202389 and a Teaching Assistantship from the Department of Electrical and Computer Engineering at University of Illinois Urbana-Champaign.
\end{acknowledgments}

%

\end{document}